\documentclass[twocolumn]{autart}    

\usepackage{graphicx}          
\usepackage{cite}
\usepackage{bm}
\usepackage{amsmath,amssymb}
\usepackage{mathrsfs}
\usepackage{amsfonts}
\usepackage{color}

\begin{document}

\begin{frontmatter}

\title{Attraction Domain Analysis for Steady States of Markovian Open Quantum Systems\thanksref{footnoteinfo}} 

\thanks[footnoteinfo]{This paper was not presented at any IFAC 
meeting. Corresponding author: Guofeng Zhang.}

\author[HK]{Shikun Zhang}\ead{shikun.zhang@polyu.edu.hk},    
\author[HK,SZ]{Guofeng Zhang}\ead{guofeng.zhang@polyu.edu.hk}           

\address[HK]{Department of Applied Mathematics, The Hong Kong Polytechnic University, Hung Hom, Kowloon, Hong Kong Special Administrative Region of China, China}  
\address[SZ]{Shenzhen Research Institute, The Hong Kong Polytechnic University, Shenzhen 518057, China}
          
\begin{keyword}                           
 open quantum systems; Lindblad master equations; steady state; attraction domain.
\end{keyword}                             

\begin{abstract}                          
This article concerns the attraction domain analysis for steady states in Markovian open quantum systems, which are mathematically described by Lindblad master equations. The central question is proposed as: given a steady state, which part of the state space of density operators does it attract and which part does it not attract? We answer this question by presenting necessary and sufficient conditions that determine, for any steady state and initial state, whether the latter belongs to the attraction domain of the former. Furthermore, it is found that the attraction domain of a steady state is the intersection between the set of density operators and an affine space which contains that steady state. Moreover, we show that steady states without uniqueness in the set of density operators have attraction domains with measure zero under some translation invariant and locally finite measures. Finally, an example regarding an open Heisenberg XXZ spin chain is presented. We pick two of the system's steady states with different magnetization profiles and analyse their attraction domains.
\end{abstract}

\end{frontmatter}

\section{Introduction}
Open quantum systems underpin the study of dissipative quantum information processing \cite{doi:10.1126/science.aaa2085}, where engineered interactions between the system and its environment facilitate various tasks including quantum metrology \cite{PhysRevResearch.2.023418}, state stabilization \cite{dissipatively,PhysRevA.78.042307,PhysRevLett.116.240503} and autonomous error correction \cite{derror,7723887}. It is thus of both theoretical and practical significance to analyse the dynamical properties of these systems. 

Under the Markov assumption, open quantum systems are mathematically described by Lindblad master equations \cite{bookoq} which induce completely positive and trace preserving dynamics on the set of density matrices. In terms of Lindblad master equations, it has been shown in \cite{PhysRevA.81.062306} that the uniqueness of a steady state is equivalent to its global attractivity, i.e., all system trajectories with initial states as density matrices converge to this state following Lindblad evolution. This important theoretical result lays the foundation for tasks related to quantum state stabilization \cite{PhysRevA.78.042307,TICOZZI20092002,PhysRevA.89.022327}. 

However, if more than one steady states are present in an open quantum system, then apparently none of them are globally attractive. It is then natural to further investigate what kind of initial states will be attracted to them and what will not, i.e., to characterize their domain of attraction. We believe that the understanding of such locally attractive dynamics may expand our knowledge on many-body quantum systems and potentially breed new mechanisms of dissipative quantum information processing. 

Characterization of attraction domains has been a topic of intensive research in classical control theory. Since exact solutions are unattainable in many cases, Lyapunov techniques are often adopted to estimate the domain of attraction \cite{8610152,ZAREI2018164,8267231}. Such methods yield inherently sufficient conditions on whether a given state belongs to the attractive region and thus result in finding subsets of exact attraction domains. In this article, however, we seek to pin down the attraction domains of steady states in open quantum systems in their entirety. To achieve this, a necessary and sufficient condition is presented to verify whether a given initial density matrix is contained in the attraction domain of a given stationary state. Apart from pointwise verification, we also characterize the global structure of attraction domains. 

Attraction domains of steady states without uniqueness are clearly proper subsets of the entire state space. Then we ask: how much ``room" in the latter does the former occupy? Measure theory is a suitable tool for answering this question. We prove that the attraction domains of nonunique steady states have measure zero under some translation invariant and locally finite measures, while there exists such a measure under which the entire state space has a positive finite measure. This sheds light on the ``almost impossibility" of stabilizing certain quantum states with Lindblad dynamics when nonlocal resources are not adequate to make them globally attractive.

\section{Preliminaries}
Let $\mathcal{H}$ be a finite dimensional Hilbert space isomorphic to $\mathbb{C}^N$, and $\mathcal{B}(\mathcal{H})$ be the set of linear operators on $\mathcal{H}$. Following Dirac's notation in quantum mechanics, the orthonormal basis of $\mathcal{H}$ is written as $\{|\epsilon_i \rangle\}_{i=1}^{N}$. The set of density operators $\mathcal{D}(\mathcal{H}) \subset \mathcal{B}(\mathcal{H})$ includes all positive semidefinite, trace-one operators on $\mathcal{H}$, which constitutes the entire state space for finite dimensional open quantum systems. Let $\rho \in \mathcal{D}(\mathcal{H})$ denote a quantum state. Its evolution according to Lindblad master equation is expressed as:
\begin{equation}\label{mainsystem}
    \dot \rho=-\text{i}[H,\rho]+\sum_{k=1}^M (L_k \rho L_k^\dagger -\frac{1}{2}L_k^\dagger L_k\rho-\frac{1}{2}\rho L_k^\dagger L_k),
\end{equation}
where Hermitian operator $H \in \mathcal{B}(\mathcal{H})$ stands for system Hamiltonian, and $L_k \in \mathcal{B}(\mathcal{H}), k=1,2,...,M$ represent coupling operators between the system and its environment. Eq. (1) can be concisely expressed in superoperator form: $\dot \rho=\mathcal{L}_{[H;L_1,...,L_M]}(\rho)$, where superoperator $\mathcal{L}_{[H;L_1,...,L_M]}$ is a linear operator from $\mathcal{B}(\mathcal{H})$ to $\mathcal{B}(\mathcal{H})$. $\forall A_1, A_2 \in \mathcal{B}(\mathcal{H})$, their inner product is defined as $\langle A_1 ,A_2 \rangle \triangleq \text{tr}(A_1^\dagger A_2)$. Consequently, the norm of $\forall \sigma  \in \mathcal{B}(\mathcal{H})$ is defined as $\Vert \sigma \Vert \triangleq \sqrt{\langle \sigma ,\sigma \rangle}$. Let the adjoint of superoperators be understood with respect to such inner product. Then, the superoperator $\mathcal{L}_{[H;L_1,...,L_M]}^\dagger$ satisfies:
\begin{equation}\label{mainsystemad}
    \begin{aligned}
    &\mathcal{L}_{[H;L_1,...,L_M]}^\dagger \big(\cdot \big) \\
    &=\text{i}[H,\cdot]+\sum_{k=1}^M L_k^\dagger (\cdot) L_k -\frac{1}{2}L_k^\dagger L_k(\cdot)-\frac{1}{2}(\cdot) L_k^\dagger L_k .
    \end{aligned}
\end{equation}

We now give the definition of steady states. 

{\bfseries Definition 1.} \textit{$\rho_{\emph{ss}} \in \mathcal{D}(\mathcal{H})$ is a steady state of system (1) if $\mathcal{L}_{[H;L_1,...,L_M]}(\rho_{\emph{ss}})=0$.}

A steady state $\rho_{\text{ss}}$ induces a bipartition of $\mathcal{D}(\mathcal{H})$. Choosing any $\rho_0 \in \mathcal{D}(\mathcal{H})$ as the initial state, its time evolved state $e^{\mathcal{L}_{[H;L_1,...,L_M]}t}\rho_0$ either converges to $\rho_{\text{ss}}$ in the long time limit or it does not. This is formalised by the following definition of attraction domains.

{\bfseries Definition 2.} \textit{The attraction domain of steady state $\rho_{\emph{ss}} \in \mathcal{D}(\mathcal{H})$ w.r.t. system dynamics (1), denoted by $\emph{DoA}[\rho_{\emph{ss}}]$, is defined as:}
\begin{equation}
    \text{DoA}[\rho_{\text{ss}}]=\{\rho \in \mathcal{D}(\mathcal{H})|\lim_{t\to +\infty}e^{\mathcal{L}_{[H;L_1,...,L_M]}t}\rho=\rho_{\text{ss}} \}.
\end{equation}
It is clear from this definition that the attraction domain of any steady state is nonempty, for it includes the steady state itself at least.

For a steady state $\rho_{\text{ss}}$, it has been shown in \cite{PhysRevA.81.062306} that $\text{DoA}[\rho_{\text{ss}}]=\mathcal{D}(\mathcal{H})$ if and only if there are no other steady states in $\mathcal{D}(\mathcal{H})$. That is, the uniqueness of $\rho_{\text{ss}}$ \big{(}in $\mathcal{D}(\mathcal{H})$\big{)} is equivalent to its global attractivity \big{(}in $\mathcal{D}(\mathcal{H})$\big{)}. 

However, if there is another steady state $\rho_{\text{ss}}' \in \mathcal{D}(\mathcal{H})$, then $\rho_{\text{ss}}$ is not globally attractive since at least $\rho_{\text{ss}}'$ does not belong to $\text{DoA}[\rho_{\text{ss}}]$ according to Definition 1. Moreover, any convex combination of $\rho_{\text{ss}}$ and $\rho_{\text{ss}}'$ is also a steady state. As a matter of fact, all steady states of Lindblad master equations in $\mathcal{D}(\mathcal{H})$ form convex sets, which may contain an uncountably infinite number of elements. In light of this, the exact characterization of attraction domains of an arbitrary steady state is a nontrivial task.

It is worthwhile comparing Lindblad master equations with Markov chains. A continuous-time homogeneous Markov chain with finite state spaces must always admit stationary distributions. This is in analogy with the fact that finite-dimensional Lindblad master equations must always admit steady states. An irreducible continuous-time homogeneous Markov chain must admit a unique stationary distribution which attracts all initial distributions \cite{norris_1997}. Irreducibility of Markov chains means that all states are accessible from one another. Consider a generalization to quantum systems, where classical state $i$ corresponds to quantum state $|i\rangle\langle i|$. Irreducibility for quantum systems can be defined as: for any initial state $|i\rangle\langle i|$, and any state $|j\rangle\langle j|$, there exists a finite time at which the time-evolved state has a nonzero overlap with $|j\rangle\langle j|$ (The overlap between two quantum states $\rho$ and $\sigma$ means $\text{tr}(\rho \sigma)$ \cite{cincio2018learning}. In the case where $\rho=|i\rangle\langle i|$ and $\sigma=|j\rangle\langle j|$ are pure quantum states, the overlap between $\rho$ and $\sigma$ is expressed as: $\text{tr}(\rho \sigma)=\vert \langle i|j\rangle\vert^2$.), which can be achieved by designing suitable Hamiltonian(s) and coupling operator(s). However, it is possible that the resulting Lindblad master equation admits more than one steady states, in which case the attraction domain of each steady state can be described by the results in our paper.

\section{Main Results}
In this section, we answer the central question of this article: if system (1) does not admit a unique steady state in $\mathcal{D}(\mathcal{H})$, then which part of $\mathcal{D}(\mathcal{H})$ does each steady state attract?

Mathematically, depicting the attraction domain of a steady state $\rho_{\text{ss}}$ is equivalent to proposing a necessary and sufficient condition that is able to verify whether any given state in $\mathcal{D}(\mathcal{H})$ belongs to $\text{DoA}[\rho_{\text{ss}}]$ or not. We thus present such a condition as one of the theoretical results of this article.


\begin{thm}
Let $\{\omega_k\}_{k=1}^J$ be a complete set of eigen-operator(s) of $\mathcal{L}_{[H;L_1,...,L_M]}^\dagger$ corresponding to eigenvalue(s) with zero real part(s). Let $\rho_{\emph{ss}}$ be an arbitrary steady state of system (1). Then, for $\rho_0 \in \mathcal{D}(\mathcal{H})$, $\rho_0 \in \emph{DoA}[\rho_{\emph{ss}}]$ if and only if $\emph{tr}(\omega_l^\dagger \rho_0)=\emph{tr}(\omega_l^\dagger \rho_{\emph{ss}})$, $1\leq l \leq J$.
\end{thm}

\begin{pf}
Let us denote the eigen-operator(s) of $\mathcal{L}_{[H;L_1,...,L_M]}^\dagger$ (\ref{mainsystemad}) with zero eigenvalue(s) by $\omega_1,...,\omega_{J_0}$ and that with purely imaginary but nonzero eigenvalues, should they exist, by $\omega_{J_0+1},...,\omega_J$. 

Superoperator $\mathcal{L}_{[H;L_1,...,L_M]}$ admits no eigenvalues with positive real parts and no generalized eigen-operators corresponding to eigenvalues with zero real parts (otherwise leading to unbounded state trajectory), and the same goes for $\mathcal{L}_{[H;L_1,...,L_M]}^\dagger$. Therefore, there exist generalized eigen-operators of $\mathcal{L}_{[H;L_1,...,L_M]}^\dagger$, denoted as $\omega_{J+1},...,\omega_{N^2}$, such that $\omega_1,...,\omega_{N^2}$ form a complete basis of $\mathcal{B}(\mathcal{H})$. 

Meanwhile, there exist $\sigma_1,...,\sigma_{N^2} \in \mathcal{B}(\mathcal{H})$, where $\{\sigma_k\}_{k=1}^{J_0}$ corresponds to eigen-operator(s) of $\mathcal{L}_{[H;L_1,...,L_M]}$ with eigenvalue 0, and $\{\sigma_k\}_{k=J_0+1}^{J}$ corresponds to eigen-operators of $\mathcal{L}_{[H;L_1,...,L_M]}$ with purely imaginary but nonzero eigenvalues, and $\{\sigma_k\}_{k=J+1}^{N^2}$ corresponds to eigen-operators and generalized eigen-operators of $\mathcal{L}_{[H;L_1,...,L_M]}$ associated with eigenvalues on the open left complex plane. The operators $\sigma_1,...,\sigma_{N^2}$ constitute a complete basis of $\mathcal{B}(\mathcal{H})$ and satisfy the following relations:
\begin{equation}
    \text{tr}(\omega_j^\dagger \sigma_j)=\delta_{ij},\quad 1\leq i,j \leq N^2 .
\end{equation}

As a result, any $\rho_0 \in \mathcal{D}(\mathcal{H})$ admits the following expansion:
\begin{equation}\label{expansion}
    \rho_0=\sum_{j=1}^{N^2}\text{tr}(\omega_j^\dagger \rho_0)\sigma_j .
\end{equation}

In terms of steady state $\rho_{\text{ss}}$, since $\mathcal{L}_{[H;L_1,...,L_M]}(\rho_{\text{ss}})=0$, it must hold that
\begin{equation}\label{zero}
   \text{tr}(\omega_i^\dagger \rho_{\text{ss}})=0,\quad J_0+1 \leq i \leq N^2  .
\end{equation}

Next, we further endow $\{\sigma_i\}_{i=J+1}^{N^2}$ and $\{\omega_i\}_{i=J+1}^{N^2}$with the following order:
\begin{equation}\label{order}
    \begin{aligned}
     &\sigma_1^{\lambda_1},...,\sigma_{n_1}^{\lambda_1};\cdots;\sigma_1^{\lambda_{m}},...,\sigma_{n_{m}}^{\lambda_{m}}\\
     &\omega_1^{\bar{\lambda}_1},...,\omega_{n_1}^{\bar{\lambda}_1};\cdots;\omega_1^{\bar{\lambda}_m},...,\omega_{n_{m}}^{\bar{\lambda}_m},
\end{aligned}
\end{equation}
where $\sigma_1^{\lambda_j}$ and $\omega_1^{\bar{\lambda}_j}$, $1 \leq j \leq m$, are eigen-operators of $\mathcal{L}_{[H;L_1,...,L_M]}$ and $\mathcal{L}_{[H;L_1,...,L_M]}^\dagger$ with eigenvalues $\lambda_j$ and $\bar{\lambda}_j$, respectively; $\sigma_i^{\lambda_j}$ and $\omega_i^{\bar{\lambda}_j}$,$1<i \leq n_j$, $1 \leq j \leq m$ are generalized eigen-operators of $\mathcal{L}_{[H;L_1,...,L_M]}$ and $\mathcal{L}_{[H;L_1,...,L_M]}^\dagger$ corresponding to eigenvalues $\lambda_j$ and $\bar{\lambda}_j$, respectively. Also, $\sum_{k=1}^m =N^2-J$.
 
Based on (\ref{expansion}) and (\ref{order}), the state trajectory starting from $\rho_0$ with $t \geq 0$ is expressed as:
\begin{equation}\label{evolution}
    \begin{aligned}
    &e^{\mathcal{L}_{[H;L_1,...,L_M]}t}\rho_0\\
    &=\sum_{i=1}^{J_0}\text{tr}(\omega_i^\dagger \rho_0)\sigma_i+\sum_{i=J_0+1}^{J}e^{\text{i}\beta_i t}\text{tr}(\omega_i^\dagger \rho_0)\sigma_i\\
    &+\sum_{k=1}^m\!  \bigg{[}\!(x_{\rho_0}^{1,\lambda_k}\!+\!x_{\rho_0}^{2,\lambda_k}t\!+\!\frac{1}{2!}x_{\rho_0}^{3,\lambda_k}t^2\!+\!\cdots \!+\!\frac{1}{(n_k\!-\!1)!}x_{\rho_0}^{n_k,\lambda_k}t^{n_k-1}) \sigma_1^{\lambda_k}\\ 
    &+(x_{\rho_0}^{2,\lambda_k}+x_{\rho_0}^{3,\lambda_k}t+\cdots+\frac{1}{(n_k-2)!}x_{\rho_0}^{n_k,\lambda_k}t^{n_k-2})\sigma_2^{\lambda_k} \\
    &+\cdots  \\ 
    &+(x_{\rho_0}^{n_k,\lambda_k})\sigma_{n_k}^{\lambda_k}\bigg{]}e^{\lambda_k t},
    \end{aligned}
\end{equation}
where 
\begin{equation}
  \mathcal{L}_{[H;L_1,...,L_M]}\sigma_i=\text{i}\beta_i \sigma_i,\quad \beta_i \in \mathbb{R},\quad J_0+1 \leq i \leq J,
\end{equation}
and for $1\leq j \leq n_k, 1 \leq k \leq m$,
\begin{equation}
    x_{\rho_0}^{j,\lambda_k}=\text{tr}\big((\omega_j^{\bar{\lambda}_k})^\dagger \rho_0 \big).
\end{equation}

Suppose that $\rho_0 \in \text{DoA}[\rho_{\text{ss}}]$. Since $\text{Re}(\lambda_k)<0$, $1\leq k \leq m$, it should hold that
\begin{equation*}
  \text{tr}(\omega_i^\dagger \rho_{\text{ss}})=\text{tr}(\omega_i^\dagger \rho_0)=0,\quad J_0+1 \leq i \leq J.
\end{equation*}
Otherwise, $e^{\mathcal{L}_{[H;L_1,...,L_M]}t}\rho_0$ does not have a limit as $t$ tends to infinity. It thus follows that
\begin{equation*}
    \lim_{t \to +\infty}e^{\mathcal{L}_{[H;L_1,...,L_M]}t}\rho_0=\sum_{i=1}^{J_0}\text{tr}(\omega_i^\dagger \rho_0)\sigma_i=\rho_{\text{ss}}.
\end{equation*}
The linear independence of $\sigma_1,...,\sigma_{J_0}$ indicates that 
\begin{equation*}
  \text{tr}(\omega_i^\dagger \rho_{\text{ss}})=\text{tr}(\omega_i^\dagger \rho_0)=0,\quad 1 \leq i \leq J_0.
\end{equation*}
Necessity is thus proved.

Next, suppose that 
\begin{equation*}
  \text{tr}(\omega_i^\dagger \rho_{\text{ss}})=\text{tr}(\omega_i^\dagger \rho_0)=0,\quad 1 \leq i \leq J.
\end{equation*}
It is clear from (\ref{zero}) and (\ref{evolution}) that 
\begin{equation*}
    \lim_{t \to +\infty}e^{\mathcal{L}_{[H;L_1,...,L_M]}t}\rho_0=\rho_{\text{ss}},
\end{equation*}
which completes the proof of sufficiency. $\hfill\square$
\end{pf}

We then present a physical interpretation of Theorem 1. In fact, based on Theorem 1, it is even possible to show that $\rho_0 \in \rho_{\text{ss}}$ if and only if there exists a linearly independent set, denoted as $\{\tilde{\omega}_k\}_{k=1}^J$, of Hermitian operator(s), such that $\text{tr}(\tilde{\omega}_k \rho_0)=\text{tr}(\tilde{\omega}_k \rho_{\text{ss}})$, $1 \leq k \leq J$. The proof is omitted for the sake of brevity. On one hand, the operator(s) in $\{\tilde{\omega}_k\}_{k=1}^J$ are Hermitian, and are thus viewed as ``observables" in quantum mechanics. On the other hand, it is clear that each observable in $\{\tilde{\omega}_k\}_{k=1}^J$ belongs to the sum of eigenspace(s) of $\mathcal{L}_{[H;L_1,...,L_M]}^\dagger$ corresponding to eigenvalue(s) with zero real part(s).

We note that the adjoint equation of (1): 
\begin{equation}
    \dot X=\mathcal{L}_{[H;L_1,...,L_M]}^\dagger (X),
\end{equation}
describes the evolution of observables in the Heisenberg picture. Therefore, the Heisenberg evolution of $\tilde{\omega}_{k_0}$, $k_0 \in \{1,...,J\}$, either remains constant (in this case, $\tilde{\omega}_{k_0}$ is called a ``conserved quantity" in \cite{PhysRevA.89.022118}) or oscillates, both displaying a non-decaying pattern in the long time limit. The $J$ non-decaying observable(s) pin down an ``identification vector" for each $\rho \in \mathcal{D}(\mathcal{H})$, which contains the expectation value(s) of the observables under state $\rho$. Theorem 1 and Proposition 1 indicate that the attraction domain of a steady state $\rho_{\text{ss}}$ is formed by the density operator(s) equipped with the same ``identification vector" as that equipped by $\rho_{\text{ss}}$.

We then present another result based on Theorem 1, which captures the global structure of attraction domains.

{\bfseries Proposition 1} \textit{Consider $\sigma_1,...,\sigma_{N^2} \in \mathcal{B}(\mathcal{H})$, where $\{\sigma_k\}_{k=1}^{J}$ corresponds to eigen-operator(s) of $\mathcal{L}_{[H;L_1,...,L_M]}$ with eigenvalue(s) admitting zero real part(s), and $\{\sigma_k\}_{k=J+1}^{N^2}$ corresponds to eigen-operator(s) and generalized eigen-operator(s) of $\mathcal{L}_{[H;L_1,...,L_M]}$ associated with eigenvalues admitting negative real parts. Let $\rho_{\emph{ss}}$ be a density operator which satisfies $\mathcal{L}_{[H;L_1,...,L_M]}(\rho_{\emph{ss}})=0$. Denote the following set:
\begin{equation}\label{apply} 
   \{\sigma \in \mathcal{B}(\mathcal{H}) | \sigma=\rho_{\emph{ss}}+\sum_{k=J+1}^{N^2}g_k \sigma_k,g_k \in \mathbb{C}, J+1 \leq k \leq N^2\}
\end{equation}
as $\mathcal{A}_{\rho_{\emph{ss}}}$, which is an affine space over the subspace of $\mathcal{B}(\mathcal{H})$ spanned by $\sigma_{J+1},...,\sigma_{N^2}$. Then, 
\begin{equation} \label{apply1}
\emph{DoA}[\rho_{\emph{ss}}]=\mathcal{A}_{\rho_{\emph{ss}}} \cap \mathcal{D}(\mathcal{H}).
\end{equation}}

\begin{pf}
We first prove that $\text{DoA}[\rho_{\text{ss}}]\subset \mathcal{A}_{\rho_{\text{ss}}} \cap \mathcal{D}(\mathcal{H})$. It suffices to prove that $\forall \rho_0 \in \text{DoA}[\rho_{\text{ss}}]$, $\rho_0 \in \mathcal{A}_{\rho_{\text{ss}}}$. 

Let $\{\eta_j\}_{j=1}^{N^2}$ be a complete set of eigen-operators and generalized eigen-operators of $\mathcal{L}_{[H;L_1,...,L_M]}^\dagger$ satisfying $\text{tr}(\eta_i^\dagger \sigma_j)=\delta_{ij}$, $1\leq i,j \leq N^2$. An arbitrary $\rho_0 \in \text{DoA}[\rho_{\text{ss}}]$ can be expanded as:
\begin{equation} 
\rho_0=\sum_{j=1}^{J}\text{tr}(\eta_j^\dagger \rho_0) \sigma_j+\sum_{j=J+1}^{N^2}\text{tr}(\eta_j^\dagger \rho_0) \sigma_j.
\end{equation}
Since $\rho_0 \in \text{DoA}[\rho_{\text{ss}}]$, according to Theorem 1, $\text{tr}(\eta_j^\dagger \rho_0)=\text{tr}(\eta_j^\dagger \rho_{\text{ss}})$, $1 \leq j \leq J$. Moreover, because
\begin{equation*}
    \rho_{\text{ss}}=\sum_{j=1}^{J}\text{tr}(\eta_j^\dagger \rho_{\text{ss}})\sigma_j,
\end{equation*}
we have
\begin{equation} 
\rho_0=\rho_{\text{ss}}+\sum_{j=J+1}^{N^2}\text{tr}(\eta_j^\dagger \rho_0) \sigma_j.
\end{equation}
which implies that $\rho_0 \in \mathcal{A}_{\rho_{\text{ss}}}$. 

Then, it shall be proved that $\mathcal{A}_{\rho_{\text{ss}}} \cap \mathcal{D}(\mathcal{H}) \subset \text{DoA}[\rho_{\text{ss}}]$. $\forall \rho_0 \in \mathcal{A}_{\rho_{\text{ss}}} \cap \mathcal{D}(\mathcal{H})$, we have
\begin{equation} 
\rho_0=\rho_{\text{ss}}+\sum_{j=J+1}^{N^2}g_{\rho_0}^j \sigma_j,  g_{\rho_0}^j \in \mathbb{C}, J+1 \leq j \leq N^2.
\end{equation}
Following (\ref{evolution}), it holds that 
\begin{equation} 
\lim_{t \to +\infty} e^{\mathcal{L}_{[H;L_1,...,L_M]}t}(\sum_{j=J+1}^{N^2}g_{\rho_0}^j \sigma_j)=0.
\end{equation}
Because $\rho_{\text{ss}}$ is a steady state, we have $\lim_{t \to +\infty} e^{\mathcal{L}t}\rho_0=\rho_{\text{ss}}$, which says that $\rho_0 \in \text{DoA}[\rho_{\text{ss}}]$. $\hfill\square$
\end{pf}

If a steady state of system (1) is not unique in $\mathcal{D}(\mathcal{H})$, its attraction domain is a strict subset of $\mathcal{D}(\mathcal{H})$. Then, how much ``volume" does this attraction domain occupy in the set of all density operators? We show that the answer is 0 under certain measures in this section. Also presented is an implication in the context of quantum state stabilization.

Before presenting the main results of this section, we make a few notations and definitions. Let us denote the set of Hermitian and trace-one operators in $\mathcal{B}(\mathcal{H})$ as $\mathcal{D}_1 (\mathcal{H})$. A subset $S$ of $\mathcal{D}_1 (\mathcal{H})$ is defined as an open set of $\mathcal{D}_1 (\mathcal{H})$ if $\forall x \in S$, there exists $\epsilon >0$, such that any $y \in \mathcal{D}_1 (\mathcal{H})$ which satisfies $\Vert x-y \Vert<\epsilon$ belongs to $S$. Let us denote the set of all open sets of $\mathcal{D}_1 (\mathcal{H})$ as $\mathcal{T}$. Also, let us denote the set of steady states of system (1) as $\Xi_{\text{ss}}$.

We are now in the position to present the following result.
\begin{thm}
Suppose that system (1) admits non-unique steady states. For any measure space of the  form $(\mathcal{D}_1 (\mathcal{H}), \Sigma,\mathcal{M})$, with $\mathcal{T} \subseteq \Sigma$, $\mathcal{M}$ being translation invariant on $\mathcal{D}_1 (\mathcal{H})$ and locally finite with respect to $(\mathcal{D}_1 (\mathcal{H}),\mathcal{T})$, and $\emph{DoA}[\rho_{\emph{ss}}]$ being measurable $\forall \rho_{\emph{ss}} \in \Xi_{\emph{ss}}$, it must hold that $\mathcal{M} (\emph{DoA}[\rho_{\emph{ss}}])=0$, $\forall \rho_{\emph{ss}} \in \Xi_{\emph{ss}}$.
\end{thm}

\begin{pf}
We shall prove Theorem 2 by contradiction. Suppose that there exists $\rho_{\text{ss}}^0 \in \Xi_{\text{ss}}$, such that $\mathcal{M}\big(\text{DoA}[\rho_{\text{ss}}^0]\big) \neq 0$. 

It is clear that $\mathcal{D}(\mathcal{H})$ is a bounded set since $\forall \rho \in \mathcal{D}(\mathcal{H})$ satisfies $\text{tr}(\rho^2) \leq 1$. Also, $\mathcal{D}(\mathcal{H})$ is closed under topology $\mathcal{T}$. Based on eqs. (\ref{apply}) and (\ref{apply1}), it also holds that $\text{DoA}[\rho_{\text{ss}}^0]$ is bounded and closed under topology $\mathcal{T}$.

Since system (1) admits non-unique steady states, there exists $\rho_{\text{ss}}^1 \in \Xi_{\text{ss}}$ which is linearly independent with $\rho_{\text{ss}}^0$. Consider the following set:
\begin{equation}\label{stra}
    \text{S}_{\text{tra}}\triangleq \bigcup_{p \in [0,1]} \big(\text{DoA}[\rho_{\text{ss}}^0]+p\{\rho_{\text{ss}}^1-\rho_{\text{ss}}^0\}\big).
\end{equation}
That is, $x \in \text{S}_{\text{tra}}$ if and only if there exists $x_0 \in \text{DoA}[\rho_{\text{ss}}^0]$ and $p \in [0,1]$, such that
\begin{equation}\label{bounded}
    x=x_0+p(\rho_{\text{ss}}^1-\rho_{\text{ss}}^0).
\end{equation}
From (\ref{bounded}), it is clear that $\text{S}_{\text{tra}}$ is bounded. We then show that $\text{S}_{\text{tra}}$ is also closed. Consider an arbitrary sequence $\{x_n\}_{n=1}^{+\infty} \subset \text{S}_{\text{tra}}$ with $\lim_{n \to +\infty}x_n=\tilde{x}$. Each $x_n$ in the sequence is decomposed as:
\begin{equation}\label{converge}
    x_n=\rho_{\text{ss}}^0+\sum_{r=J+1}^{N^2}(g_r^n)\sigma_r+\tilde{p}_n (\rho_{\text{ss}}^1-\rho_{\text{ss}}^0),
\end{equation}
where $n \geq 1$, $\{g_r^n\}_{n=1}^{+\infty}\subset \mathbb{C}$ ($J+1\leq r \leq N^2$), $\tilde{p}_n \in [0,1]$, and $\{\sigma_r\}_{r=J+1}^{N^2}$ is defined in Proposition 1. Since the sequence $\{x_n\}_{n=1}^{+\infty}$ is convergent, it must be a Cauchy sequence. Therefore, it holds that, for $k>m$
\begin{equation}
    \lim_{m,k \to +\infty} \sum_{r=J+1}^{N^2}(g_r^k-g_r^m)\sigma_r+(\tilde{p}_{k}-\tilde{p}_{m})(\rho_{\text{ss}}^1-\rho_{\text{ss}}^0)=0.
\end{equation}
Because $\sigma_{J+1},...,\sigma_{N^2}$ and $\rho_{\text{ss}}^1-\rho_{\text{ss}}^0$ are linearly independent, it follows that $\{g_r^n\}_{n=1}^{+\infty}$ ($J+1 \leq r \leq N^2$) and $\{\tilde{p}_{n}\}_{n=1}^{+\infty}$ are all Cauchy sequences and are therefore convergent. Consequently, on one hand, the sequence $\{\rho_{\text{ss}}^0+\sum_{r=J+1}^{N^2}(g_r^n)\sigma_r\}_{n=1}^{+\infty}$ converges, and since each element in the sequence belongs to the closed set $\text{DoA}[\rho_{\text{ss}}^0]$, its limit also belongs to $\text{DoA}[\rho_{\text{ss}}^0]$. On the other hand, the sequence $\{\tilde{p}_{n}\}_{n=1}^{+\infty}$ converges to a limit in $[0,1]$, since $[0,1]$ is a closed set. Therefore, we have shown that $\tilde{x} \in \text{S}_{\text{tra}}$, which says that $\text{S}_{\text{tra}}$ is closed.

Since $\text{S}_{\text{tra}}$ (\ref{stra}) is bounded and closed, it is a compact set. The fact that $\mathcal{M}$ is locally finite indicates that $0 \leq \mathcal{M}(\text{S}_{\text{tra}})<+\infty$. Consider a infinite sequence $\{p_n\}_{n=1}^{+\infty} \subset [0,1]$ with $p_i \neq p_j$, $i,j \geq 1$. Because the $\mathcal{M}$ is translation invariant, it holds that
\begin{equation*}
    \mathcal{M}(\text{DoA}[\rho_{\text{ss}}^0])=\mathcal{M}\big(\text{DoA}[\rho_{\text{ss}}^0]+p_n\{\rho_{\text{ss}}^1-\rho_{\text{ss}}^0\}\big),
\end{equation*}
where $n \geq 1$. Based on Proposition 1, the sets $\text{DoA}[\rho_{\text{ss}}^0]+p_n\{\rho_{\text{ss}}^1-\rho_{\text{ss}}^0\}\triangleq S_n$ ($n \geq 1$) are mutually disjoint. The countable additivity and monotonicity of measure $\mathcal{M}$ say that
\begin{equation}
    \mathcal{M}(\bigcup_{n=1}^{+\infty}S_n)\!=\!\sum_{n=1}^{+\infty}\mathcal{M}(S_n)\!=\!\sum_{n=1}^{+\infty}\mathcal{M}(\text{DoA}[\rho_{\text{ss}}^0])\!\leq \!\mathcal{M}(\text{S}_{\text{tra}}).
\end{equation}
If $\mathcal{M}(\text{DoA}[\rho_{\text{ss}}^0]) \neq 0$, then $\mathcal{M}(\text{S}_{\text{tra}})$ cannot be finite, which leads to a contradiction. Therefore, $\mathcal{M}(\text{DoA}[\rho_{\text{ss}}^0])=0$. $\hfill\square$
\end{pf}
Next, we show in the following theorem that it is possible to construct a translation invariant and locally finite measure, under which $\mathcal{D}(\mathcal{H})$ has a finite positive measure, while attraction domains of non-unique steady states have measure zero.

\begin{thm}
There exists a measure space $\big(\mathcal{D}_1 (\mathcal{H}),\Sigma_0, \mathcal{M}_0 \big)$, where $\mathcal{T}\subseteq \Sigma_0$, and $\mathcal{M}_0$ is translation invariant on $\mathcal{D}_1 (\mathcal{H})$ and locally finite with respect to $(\mathcal{D}_1 (\mathcal{H}),\mathcal{T})$. With this measure space, $\mathcal{D}(\mathcal{H})$ is measurable and $0<\mathcal{M}_0 \big(\mathcal{D}(\mathcal{H})\big)< +\infty$. Also, $\emph{DoA}[\rho_{\emph{ss}}]$ is measurable $\forall \rho_{\emph{ss}} \in \Xi_{\emph{ss}}$, and $\mathcal{M}_0 (\emph{DoA}[\rho_{\emph{ss}}])=0$ if $\rho_{\emph{ss}}$ is not the unique steady state in $\Xi_{\emph{ss}}$.
\end{thm}

\begin{pf}
$\forall A \in \mathcal{D}_1 (\mathcal{H})$, there exist $a_1,...,a_{N^2-1} \in \mathbb{R}$, such that 
\begin{equation}
    A=\frac{1}{N} I_N+\sum_{k=1}^{N^2-1}\frac{a_k}{\sqrt{2}}B_k,
\end{equation}
where $B_1,...,B_{N^2-1}$ are generalized Gell-Mann matrices \cite{6580216}. Therefore, there exists a bijection $f$ between $\mathcal{D}_1 (\mathcal{H})$ and $\mathbb{R}^{N^2-1}$ mapping $A$ to $(a_1,...,a_{N^2-1})$. Consequently, there is also a bijection $\mathcal{F}$ between $2^{\mathcal{D}_1 (\mathcal{H})}$ and $2^{\mathbb{R}^{N^2-1}}$ which satisfies
\begin{equation*}
    \mathcal{F}(\Omega)=\{y\in \mathbb{R}^{N^2-1}|\exists x \in \Omega, f(x)=y\}, \forall \Omega \in 2^{\mathcal{D}_1 (\mathcal{H})}.
\end{equation*}
Denote the set of all Lebesgue measurable sets in $2^{\mathbb{R}^{N^2-1}}$ as $\Sigma_L$. Next, we define the set of all measurable sets in $2^{\mathcal{D}_1 (\mathcal{H})}$ as $\mathcal{F}^{-1}(\Sigma_L)\triangleq \Sigma_0$. Then, the measure $\mathcal{M}_0$ is defined by 
\begin{equation}\label{m0}
    \mathcal{M}_0(\Omega)\triangleq m\big(\mathcal{F}(\Omega)\big), \quad \Omega \in \Sigma_0,
\end{equation}
where $m$ is the Lebesgue measure on $\mathbb{R}^{N^2-1}$. 

Consider a translation $\sigma_{\text{tra}}$ on $\mathcal{D}_1 (\mathcal{H})$, i.e., $\forall \rho \in \mathcal{D}_1 (\mathcal{H})$, $\rho+\sigma_{\text{tra}} \in \mathcal{D}_1 (\mathcal{H})$. Then, there exists a unique $(\sigma_{\text{tra}}^1,...,\sigma_{\text{tra}}^{N^2-1})\triangleq \sigma_{\text{tra}}^{\text{vec}} \in \mathbb{R}^{N^2-1}$, such that if 
\begin{equation*}
    \rho=\frac{1}{N} I_N+\sum_{k=1}^{N^2-1}\frac{b_k}{\sqrt{2}}B_k,
\end{equation*}
then
\begin{equation*}
    \rho+\sigma_{\text{tra}}=\frac{1}{N} I_N+\sum_{k=1}^{N^2-1}\frac{b_k+\sigma_{\text{tra}}^{k}}{\sqrt{2}}B_k,
\end{equation*}
$\forall \rho \in \mathcal{D}_1 (\mathcal{H})$. As a result, $\forall \Omega \in \Sigma_0$,
\begin{equation}
    \mathcal{F}(\Omega+\{\sigma_{\text{tra}}\})=\mathcal{F}(\Omega)+\{\sigma_{\text{tra}}^{\text{vec}}\}.
\end{equation}
The Lebesgue measure $m$ is translation invariant, and the set $\mathcal{F}(\Omega)+\{\sigma_{\text{tra}}^{\text{vec}}\}$ is Lebesgue measurable. We have the following equations:
\begin{equation}
\begin{aligned}
 &\mathcal{M}_0 \big(\Omega+\{\sigma_{\text{tra}}\}\big)=m\big(\mathcal{F}(\Omega+\{\sigma_{\text{tra}}\})\big)\\
 &=m\big(\mathcal{F}(\Omega)+\{\sigma_{\text{tra}}^{\text{vec}}\}\big)=m\big(\mathcal{F}(\Omega)\big)\\
 &=\mathcal{M}_0 (\Omega),
\end{aligned}
\end{equation}
which says that $\mathcal{M}_0$ (\ref{m0}) is translation invariant.

Let us denote the collection of all open sets of $\mathbb{R}^{N^2-1}$ induced by Euclidean distance as $\mathcal{T}_1$ (i.e., $T \subseteq \mathbb{R}^{N^2-1}$ is an open set \textit{iff} $\forall x \in T$, $\exists \epsilon_x >0$, the open ball $B(x,\epsilon_x) \subseteq T$). Then, the mapping $\mathcal{F}$ establishes a one-to-one correspondence between $\mathcal{T}$ and $\mathcal{T}_1$. Since $m$ is locally finite, $\mathcal{T}_1 \subseteq \Sigma_L$. Therefore, it is true that $\mathcal{T} \subseteq \Sigma_0$, and that $\mathcal{M}_0$ is locally finite.

It is clear that $\mathcal{D}(\mathcal{H})$ is a convex subset of $\mathcal{D}_1 (\mathcal{H})$, which implies that $\mathcal{F}\big(\mathcal{D}(\mathcal{H})\big) \subseteq \mathbb{R}^{N^2-1}$ is also convex. As a result, $\mathcal{F}\big(\mathcal{D}(\mathcal{H})\big)$ is Lebesgue measurable \cite{lang1986note}, and thus $\mathcal{D}(\mathcal{H})$ is measurable.  

Next, since $\frac{1}{N}I_N \in \mathcal{D}(\mathcal{H})$ is positive definite, there exists $\epsilon>0$, such that the set:
\begin{equation}
    S(\epsilon)\triangleq\{\rho|\rho=\frac{1}{N}I_N +\!\sum_{k=1}^{N^2-1}a_k B_k, |a_k|< \! \epsilon, 1\leq k \leq \! N^2-1\}
\end{equation}
is a subset of $\mathcal{D}(\mathcal{H})$. The set $\mathcal{F}\big(S(\epsilon)\big)$ is a $N^2-1$ dimensional hypercube in $\mathbb{R}^{N^2-1}$ and is thus Lebesgue measurable. Therefore, $S(\epsilon)$ is measurable and it holds that
\begin{equation}
    \mathcal{M}_0\big(S(\epsilon)\big)=m\bigg(\mathcal{F}\big(S(\epsilon)\big)\bigg)=\epsilon^{N^2-1}>0.
\end{equation}
It follows that $0<\mathcal{M}_0\big(\mathcal{D}(\mathcal{H})\big)< +\infty$ since it is compact and it contains a subset with positive measure.

Let us now consider steady state $\rho_{\text{ss}}$ which is not the unique steady state in $\mathcal{D}(\mathcal{H})$. Since $\text{DoA}[\rho_{\text{ss}}]$ and $\mathcal{F}\big(\text{DoA}[\rho_{\text{ss}}]\big)$ are both convex, $\text{DoA}[\rho_{\text{ss}}]$ is measurable. As a consequence of Theorem 4, we have $\mathcal{M}_0 \big(\text{DoA}[\rho_{\text{ss}}]\big)=0$. $\hfill\square$


\end{pf}



\section{Example}
In this section, we show an application of our work in physics. Condensed matter physics \cite{marder2010condensed} has been an integral part of physical sciences. With the advent of quantum mechanics, quantum many-body systems\cite{tasaki2020physics}, not surprisingly, has been intensively studied by condensed matter physicists. Acknowledging that quantum systems may interact with their environment, it is then difficult to ignore open quantum many-body systems, which certainly include those described by Lindblad master equations (\ref{mainsystem}). In fact, of particular interests to physicists are steady states of these open many-body systems with different kinds of physical properties, e.g., \cite{PhysRevA.91.042117,PhysRevLett.116.235302}. It is beyond the scope of this paper to investigate in detail the steady states' physical properties. What we hope, from the perspective of Systems and Control science, is to provide a generic theoretical tool, with which physicists may know what are the initial conditions needed to witness the steady states that interest them. 

As an example, we analyse the attraction domain of steady states of a Heisenberg XXZ spin chain \cite{F_Bonechi_1992} with different spin currents. Spin currents (as will be clarified later) of steady states have been considered in, e.g., \cite{PhysRevLett.106.220601,Buča_2012}.

In quantum mechanics, the Hilbert space for a two-level system is $\mathcal{H}_s\triangleq\mathbb{C}^2$. For composite systems with multiple two-level systems, the underlying Hilbert spaces are tensor products of single qubit Hilbert spaces. The following notations are made:
\begin{equation}
    |1\rangle\triangleq \begin{pmatrix}1\\0 \end{pmatrix},\quad |0\rangle\triangleq \begin{pmatrix}0\\1 \end{pmatrix},
\end{equation}

\begin{equation}\label{sigmas}
    \sigma^-\triangleq |1\rangle\langle 0|, \sigma^+\triangleq |0\rangle\langle 1|, \sigma^z \triangleq |1\rangle\langle 1|-|0\rangle\langle 0|.
\end{equation}

An open Heisenberg XXZ spin chain with a finite length $N_c$ is a one-dimensional chain of $N_c$ two-level systems. The Hilbert space for the chain is expressed as $\mathcal{H}_c \triangleq \bigotimes_{n=1}^{N_c} \mathcal{H}_s$, where ``$\bigotimes$" denotes tensor product. 
The Hamiltonian reads:
\begin{equation}
    H=\sum_{j=1}^{N_c-1} 2(\sigma_j^- \sigma_{j+1}^+ + \sigma_j^+ \sigma_{j+1}^-)+\sigma_j^z \sigma_{j+1}^z.
\end{equation}
We note that $\sigma_j^-$, $\sigma_j^+$ and $\sigma_j^z$, $1\leq j \leq N_c$, are shorthand notations for operators acting as $\sigma^-$, $\sigma^+$ and $\sigma^z$ (\ref{sigmas}) on qubit $j$, respectively, and as the identity operator on other qubits. For example, $\sigma_1^- \triangleq \sigma^- \otimes I_{2^{N_c -1}}$. This Hamiltonian models spin exchange and Z-Z type interaction. The two coupling operators are expressed as\cite{Buča_2012}:
\begin{equation}
    L_1=2\sigma_1^- \sigma_{N_c}^+,\quad L_2=\sigma_1^+ \sigma_{N_c}^- ,
\end{equation}
which model nonlocal source-and-sink effect on both ends of the chain. Moreover, the spin current operator on site $i$, $2 \leq i \leq N_c -1$ is defined as
\begin{equation}
    J_i\triangleq \sigma_i^x \sigma_{i+1}^y - \sigma_i^y \sigma_{i+1}^x.
\end{equation}
The spin current on site $i$ with state $\rho$, $2 \leq i \leq N_c -1$, is defined as $\text{tr}(\rho J_i)$ \cite{Buča_2012}, which can be viewed as an indicator of how fast magnetization is transported in the system.

We fix $N_c=4$. In this case, the kernel of $\mathcal{L}_{[H,L_1,L_2]}$ is a 10-dimensional subspace of $\mathcal{B}(\mathcal{H}_c)$, and the system admits an uncountably infinite number of steady states. In this example, we focus on two of the set of all steady states. The first one, $\rho_{\text{ss,1}}$, reads
\begin{equation}\label{steady1}
        \rho_{\text{ss,1}} \triangleq |\Psi\rangle\langle\Psi|,\quad |\Psi\rangle=\frac{1}{\sqrt{2}}(|0110\rangle-|1001\rangle).
\end{equation}
The second one, $\rho_{\text{ss,2}}$, admits no simple expression. We present an approximate expression, keeping only 5 digits for real and imaginary parts of coefficients:
\begin{multline}\label{steady2}
    0.3401E_{44}+0.2770E_{33}+0.2308E_{22}+0.1521E_{11}+\\
    \big{\{}(0.0833+0.0671\text{i})E_{43}+\big((0.0370+0.0463\text{i})E_{42}+0.0370\text{i}E_{41}\\
    +(0.0347+0.0671\text{i})E_{32}+(-0.0093+0.0463\text{i})E_{31}\\
    +(-0.0208+0.0671\text{i})E_{21}+H.c.\big{\}},
\end{multline}
where $E_{ij}\triangleq |e_i\rangle\langle e_j|$, $1\leq i,j \leq 4$, and $|e_1\rangle=|0111\rangle$, $|e_2\rangle=|1011\rangle$, $|e_3\rangle=|1101\rangle$, $|e_4\rangle=|1110\rangle$. Also, $H.c.$ means Hermitian conjugate. 

It is checked that $\text{tr}(\rho_{\text{ss,1}} J_i)=0$, while $\text{tr}(\rho_{\text{ss,2}} J_i)\approx 0.2684$, $i=2,3$, which signifies different physical properties. The nonexistence of spin currents regarding $\rho_{\text{ss,1}}$ is associated with the concept of insulators \cite{PhysRevLett.106.220601}, while $\rho_{\text{ss,2}}$ supports spin currents. Spin currents may, for example, find application in probing quantum spin liquids \cite{han2020spin}.

We then proceed to analyse the attraction domains of these two steady states. In this example, 14 linearly independent eigenoperators of $\mathcal{L}_{[H,L_1,L_2]}^\dagger$ with zero real parts are found, which leads to the following 14 linearly independent non-decaying observables $\omega_1$---$\omega_{14}$.
\begin{equation}
    \begin{aligned}
        &\omega_1=\Pi_4,\quad \omega_2=\Pi_2,\quad \omega_3=\Pi_0,\\
        &\omega_4=\Pi_{-2},\quad \omega_5=\Pi_{-4},
    \end{aligned}
\end{equation}
where $\Pi_k$ denotes the orthogonal projection onto the following subspaces $\mathcal{S}_k$, $k=4,2,0,-2,-4$:
\begin{equation}
    \begin{aligned}
        &\mathcal{S}_4\triangleq \text{span}\{|1111\rangle\}\\
        &\mathcal{S}_2\triangleq \text{span}\{|1110\rangle,|1101\rangle,|1011\rangle,|0111\rangle\}\\
        &\mathcal{S}_0\triangleq \text{span}\{|1100\rangle,|1010\rangle,|0110\rangle,|1001\rangle,|0101\rangle,|0011\rangle\}\\
        &\mathcal{S}_{-2}\triangleq \text{span}\{|1000\rangle,|0100\rangle,|0010\rangle,|0001\rangle\}\\
        &\mathcal{S}_{-4}\triangleq \text{span}\{|0000\rangle\}.
    \end{aligned}
\end{equation}

\begin{equation}
    \begin{aligned}
        &\omega_6=|1111\rangle\langle0000|+|0000\rangle\langle1111|\\
        &\omega_7=\text{i}(|1111\rangle\langle0000|-|0000\rangle\langle1111|),
    \end{aligned}
\end{equation}

\begin{multline}
\omega_8=|1110\rangle\langle1000|+|1000\rangle\langle1110|+|1101\rangle\langle0100|\\
+|0100\rangle\langle1101|+|1011\rangle\langle0010|+|0010\rangle\langle1101|\\
+|0111\rangle\langle0001|+|0001\rangle\langle0111|,
\end{multline}

\begin{multline}
\omega_9=\text{i}(|1110\rangle\langle1000|-|1000\rangle\langle1110|)+\text{i}(|1101\rangle\langle0100|\\
-|0100\rangle\langle1101|)+\text{i}(|1011\rangle\langle0010|-|0010\rangle\langle1101|)\\
+\text{i}(|0111\rangle\langle0001|-|0001\rangle\langle0111|),
\end{multline}

\begin{equation}
    \omega_{10}=(|0110\rangle-|1001\rangle)(\langle0110|-\langle1001|),
\end{equation}

\begin{equation}
    \begin{aligned}
        &\omega_{11}=|1111\rangle(\langle0110|-\langle1001|)+(|0110\rangle-|1001\rangle)\langle1111|\\
        &\omega_{12}=\text{i}(|1111\rangle(\langle0110|\!-\!\langle1001|)\!-\!(|0110\rangle\!-\!|1001\rangle)\langle1111|),
    \end{aligned}
\end{equation}

\begin{equation}
    \begin{aligned}
        &\omega_{13}=|0000\rangle(\langle0110|-\langle1001|)+(|0110\rangle-|1001\rangle)\langle0000|\\
        &\omega_{14}=\text{i}(|0000\rangle(\langle0110|\!-\!\langle1001|)\!-\!(|0110\rangle\!-\!|1001\rangle)\langle0000|).
    \end{aligned}
\end{equation}

Based on Theorem 1, an initial state $\rho_0$ belongs to the attraction domain of a steady state $\rho_{\text{ss}}$ in this example if and only if $\text{tr}(\omega_j^\dagger \rho_0)=\text{tr}(\omega_j^\dagger \rho_{\text{ss}})$, $1\leq j \leq 14$. We present an interpretation of these 14 equalities. 

(i) $1\leq j\leq 5$: A steady state $\rho_{\text{ss}}$ and any $\rho_0 \in \text{DoA}[\rho_{\text{ss}}]$ must have the same amount of projection onto subspaces $\mathcal{S}_4$, $\mathcal{S}_2$, $\mathcal{S}_0$, $\mathcal{S}_{-2}$ and $\mathcal{S}_{-4}$;

(ii) $j=6,7$: A steady state $\rho_{\text{ss}}$ and any $\rho_0 \in \text{DoA}[\rho_{\text{ss}}]$ must have the same amount of coherence between $|1111\rangle$ and $|0000\rangle$, i.e., $\langle1111| \rho_{\text{ss}}|0000\rangle=\langle 1111| \rho_0|0000\rangle$;

(iii) $j=8,9$: A steady state $\rho_{\text{ss}}$ and any $\rho_0 \in \text{DoA}[\rho_{\text{ss}}]$ must have the same amount of summed coherence between the following four pairs of states: $(|1110\rangle,|1000\rangle)$, $(|1101\rangle,|0100\rangle)$, $(|1011\rangle,|0010\rangle)$, $(|0111\rangle,|0001\rangle)$;

(iv) $j=10$: A steady state $\rho_{\text{ss}}$ and any $\rho_0 \in \text{DoA}[\rho_{\text{ss}}]$ must have the same amount of projection onto $|0110\rangle-|1001\rangle$;

(v) $j=11,12$: A steady state $\rho_{\text{ss}}$ and any $\rho_0 \in \text{DoA}[\rho_{\text{ss}}]$ must have the same amount of coherence between $|1111\rangle$ and $|0110\rangle-|1001\rangle$;

(vi) $j=13,14$: A steady state $\rho_{\text{ss}}$ and any $\rho_0 \in \text{DoA}[\rho_{\text{ss}}]$ must have the same amount of coherence between $|0000\rangle$ and $|0110\rangle-|1001\rangle$;

On one hand, it is at this point clear that the attraction domain of $\rho_{\text{ss,1}}$ in (\ref{steady1}) is trivial, i.e., 
\begin{equation}\label{trivial}
   \text{DoA}[\rho_{\text{ss,1}}]=\{\rho_{\text{ss,1}}\}.
\end{equation}
This is because $\rho_{\text{ss,1}}$ is the only state in $\mathcal{D}(\mathcal{H}_c)$ which is fully supported by $\text{span}\{|0110\rangle-|1001\rangle\}$. 

On the other hand, the attraction domain of $\rho_{\text{ss,2}}$ with approximate expression (\ref{steady2}) is nontrivial. In fact, it is possible to show that 
\begin{equation}\label{nontrivial}
   \text{DoA}[\rho_{\text{ss,2}}]=\mathcal{D}(\mathcal{S}_2).
\end{equation}
Since $\text{tr}(\omega_2^\dagger \rho_{\text{ss,2}})=1$, any $\rho_0 \in \text{DoA}[\rho_{\text{ss}}]$ must satisfy $\text{tr}(\omega_2^\dagger \rho_0)=1$, which says that $\rho_0 \in \mathcal{D}(\mathcal{S}_2)$. Therefore, $\text{DoA}[\rho_{\text{ss,2}}]\subseteq \mathcal{D}(\mathcal{S}_2)$. Conversely, it is checked that
\begin{equation*}
\begin{aligned}
    &\text{tr}(\omega_j^\dagger \rho_{\text{ss,2}})=\text{tr}(\omega_j^\dagger \rho_0)=0,1\leq j \leq 14, j \neq 2,\\
    &\text{tr}(\omega_2^\dagger \rho_{\text{ss,2}})=\text{tr}(\omega_2^\dagger \rho_0)=1.
\end{aligned}
\end{equation*}
for all $\rho_0 \in \mathcal{D}(\mathcal{S}_2)$. We thus arrive at $\mathcal{D}(\mathcal{S}_2)\subseteq \text{DoA}[\rho_{\text{ss,2}}]$.

Next, consider choosing the following three initial states in $\mathcal{D}(\mathcal{S}_2)$, namely,
\begin{equation}
    \begin{aligned}
        &\rho_{0,1}=|1101\rangle\langle1101|\\
        &\rho_{0,2}=\frac{1}{2}(|1110\rangle+|1011\rangle)(\langle1110|+\langle1011|)\\
        &\rho_{0,3}=\frac{1}{2}|0111\rangle\langle0111|+\frac{1}{3}|1011\rangle\langle1011|+\frac{1}{6}|1110\rangle\langle1110|.
    \end{aligned}
\end{equation}
Fig.1 shows the simulated variation of $\Vert e^{\mathcal{L}_{[H,L_1,L_2]} t}\rho_{0,j}-\rho_{\text{ss,2}}\Vert_2$, $1\leq j\leq 3$, i.e., the distance between the time-evoluted states with $\rho_{\text{ss,2}}$ measured by 2-norm, with state trajectories starting from these three initial states. It is observed that the distances approach 0, which is in accordance with the fact that these initial states belong to $\text{DoA}[\rho_{\text{ss,2}}]$.
\begin{figure}
    \centering
    \includegraphics[height=6.5cm]{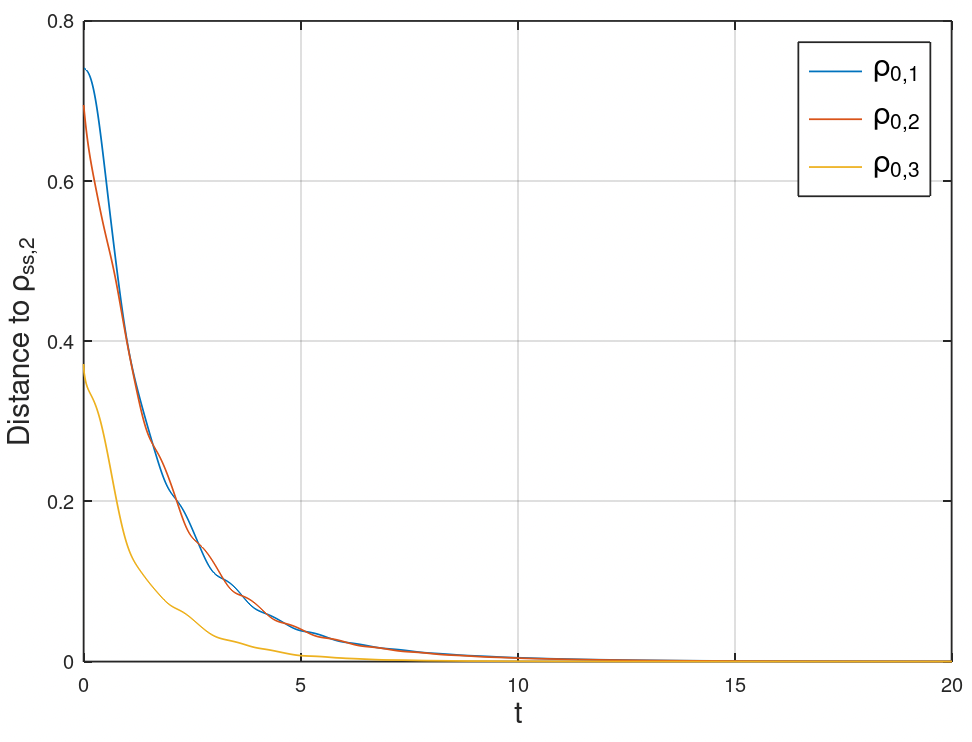}    
\caption{Simulated variation of distances between time-evoluted states with $\rho_{\text{ss,2}}$. }  
 \label{fig2}
\end{figure}

Finally, since the system admits more than one steady states, following Theorem 3, $\text{DoA}[\rho_{\text{ss,1}}]$ and $\text{DoA}[\rho_{\text{ss,2}}]$ have zero ``volume" under some translation invariant and locally finite measures.

\section{Conclusion}
We have presented an analysis on the attraction domain of steady states of finite-level quantum systems described by Lindblad master equations. Necessary and sufficient conditions are given for verification of whether an initial state belongs to the attraction domain of a steady state. We have also shown that steady states that are not unique have attraction domains with measure zero under certain measures.

\begin{ack}                               
This research is partially supported by Hong Kong Research Grant Council (Grants Nos. 15203619 and 15208418), Shenzhen Fundamental Research Fund, China, under Grant No. JCYJ20190813165207290, National Natural Science Foundation of China under Grant No. 62173269, and the CAS AMSS-polyU Joint Laboratory of Applied Mathematics.  
\end{ack}

\bibliographystyle{apalike}        
\bibliography{autosam}           

\end{document}